\documentclass[conference]{IEEEtran}
\usepackage{amsmath,amsfonts}
\usepackage{algorithmic}
\usepackage{array}
\usepackage[caption=false,font=normalsize,labelfont=sf,textfont=sf]{subfig}
\usepackage{textcomp}
\usepackage{stfloats}
\usepackage{url}
\usepackage{verbatim}
\usepackage{graphicx}
\usepackage{amssymb}
\def\BibTeX{{\rm B\kern-.05em{\sc i\kern-.025em b}\kern-.08em
    T\kern-.1667em\lower.7ex\hbox{E}\kern-.125emX}}
\usepackage{balance}

\begin{document}

\title{On Outage Analysis of OTFS Based LEO-Satellite Systems With HAPS Relaying}

\author{\IEEEauthorblockN{
Burak Ahmet Çelebi\IEEEauthorrefmark{1}\IEEEauthorrefmark{2},
Ömer Faruk Akyol\IEEEauthorrefmark{1}\IEEEauthorrefmark{2},
Semiha Tedik Başaran\IEEEauthorrefmark{2},
İbrahim Hökelek\IEEEauthorrefmark{1},
Ali Görçin\IEEEauthorrefmark{1}\IEEEauthorrefmark{2}
}
\IEEEauthorblockA{\IEEEauthorrefmark{1} Communications and Signal Processing Research (HİSAR) Lab., TÜBİTAK-BİLGEM, Kocaeli, Türkiye}
\IEEEauthorblockA{\IEEEauthorrefmark{2} Department of Electronics and Communications Engineering, Istanbul Technical University, İstanbul, Türkiye}
Emails: \{celebib16, akyolo15, tedik, aligorcin\}@itu.edu.tr,\\ibrahim.hokelek@tubitak.gov.tr}

\maketitle

\begin{abstract}
This paper presents an Orthogonal Time Frequency Space (OTFS) waveform application along with a high altitude platform station (HAPS) relaying for remedying severe Doppler effects in non-terrestrial networks (NTNs). Taking practical challenges into consideration, HAPS is exploited as a decode and forward relay node to mitigate the high path loss between a satellite and a base station (BS). In addition, a maximum ratio transmission scheme with multiple antennas at the LEO-satellite is utilized to maximize Signal-to-Noise Ratio (SNR). A shadowed Rician fading model is employed for the channel realization between the LEO-satellite and the HAPS while Nakagami-$m$ is used between the HAPS and the BS. We derive the closed-form expression of the outage probability (OP) for the end-to-end system. The theoretical and simulation results demonstrate that the {OP} can significantly decrease when the OTFS order and the number of transmit antennas increase.
\end{abstract}
\section{Introduction}
\IEEEPARstart{L}{ow-earth orbit (LEO)} satellites and high altitude platform stations (HAPSs) have become highly popular as they can provide wireless communication coverage over a large geographic area. Providing a reliable signal transmission of non-terrestrial networks (NTNs) becomes a challenging task due to harsh channel characteristics stemming from higher mobility, wireless signal propagation distance, and atmospheric conditions. Conventional orthogonal frequency-division multiplexing (OFDM), which has been heavily utilized for wireless communication in existing LEO satellite deployments \cite{Fast_Tracking}, is highly sensitive to Doppler effects for fast-fading LEO channels \cite{Performance_of_frequency}. The orthogonal time frequency space (OTFS) modulation can be an effective solution by utilizing the advantages of the delay-Doppler (DD) domain, particularly under high-mobility scenarios~\cite{Delay-Doppler}. {The performance comparison of the OTFS and OFDM waveforms is investigated in terms of outage, bit error rate, and achievable capacity~\cite{Comparison1_OTFSPerformanceonStaticMultipathChannels,Outage,Comparison2_Rate}}. 

 LEO-satellite communication systems such as Starlink and OneWeb \cite{Starlink} have been increasingly deployed to provide commercial services including navigation, broadcasting, and communication. The quality of these services highly depends on the wireless propagation channel between the LEO-satellite and the ground station \cite{New_Model}. Large-scale \cite{Large_Scale2} and small-scale \cite{New_Model} fading characteristics of NTN wireless channel have been extensively investigated in the literature. Among these studies, the analytical model characterizing the LEO-satellite channel \cite{New_Model} has been adapted by many studies due to its tractability and high accuracy with experimental measurements.

Using a relay node to solve the non-line of sight (NLOS) problem or mitigate the effect of high path-loss is a popular technique for wireless communication systems including NTNs~\cite{Haps_Based_Relaying,Dual_Hop,Dual-Hop-ref1}. For example, an integrated space-air-ground uplink network utilizing HAPS as a relay station is proposed for achieving higher reliability~\cite{Haps_Based_Relaying}. A satellite is utilized as a relay node between ground-based source and destination nodes \cite{Dual_Hop, Dual-Hop-ref1}. The closed form equations of dual-hop communication with multiple input multiple output (MIMO) systems for shadowed Rician (SR) channel is derived in \cite{Dual_Hop} for an OFDM waveform. There has been a significant effort to develop an OTFS waveform to mitigate the Doppler effect in wireless communication between LEO-satellite and ground station \cite{High_Mobility_Satellite, OTFS_Modulation_Performance}. Different channel equalization methods and realistic frame structures for a satellite-to-ground channel employing an OTFS modulation are investigated through simulation experiments in \cite{OTFS_Modulation_Performance}. None of the above studies models the outage probability ({OP}) of the dual-hop communication system with an OTFS waveform. In \cite{Outage}, zero forcing (ZF) equalization with decode-and-forward (DF) relaying is studied. Although the study derives a closed-form expression for the outage probability, the simulation results are not verified with theoritical deductions. They assume that the link between satellite and UAV nodes is ideal while our study utilizes a random channel model not only for HAPS to base station (BS) but also LEO-satellite to HAPS links.

 Our study, which is inspired from \cite{Outage}, provides analytical model for the outage performance of the OTFS-based LEO-satellite system with a dual-hop relaying. HAPS is deployed as a DF relay to mitigate high path losses while a maximum ratio transmission (MRT) scheme with multiple antennas at the LEO-satellite is utilized to enhance signal-to-noise ratio (SNR). The SR fading model is employed for the channel realization between the LEO-satellite and HAPS while Nakagami-$m$ is utilized between HAPS and BS. The OTFS waveform along with its mathematical underpinnings are first described. The use of the MRT scheme in the context of multiple input single output (MISO) channels with an OTFS waveform is then investigated by deriving a closed-form expression of the {OP} for the wireless link between the LEO-satellite and the HAPS. Finally, the performance of the MRT with ZF equalization for the SR channels is presented under both heavy and Karasawa shadowing conditions. The results show that the {OP} can be decreased as the OTFS order and the number of transmit antennas increase.

The rest of the paper is organized as follows. Section II presents the system model including the channel model and OTFS waveform. In Section III, the {OP} analysis for the LEO-satellite network with cooperative HAPS is presented. The analytical and simulation results are provided in Section IV. Finally, Section V concludes the study.

\section{System Model}
A downlink LEO-satellite communication system consisting of one LEO-satellite, a cooperative HAPS, and a BS is shown in Fig.\ref{fig:1}. The HAPS hovering in the air is utilized as a relay node to reduce the effect of path loss over the communication link between the LEO satellite and the BS. Both the HAPS and the BS are equipped with a single antenna, while the LEO satellite is equipped with {$K$} antennas. There is no direct communication link between the LEO satellite and the BS. The LEO satellite is assumed to be moving at a high speed.
\subsection{OTFS Waveform}
\begin{figure}[t!]
    \centering
    \includegraphics[width=\linewidth]{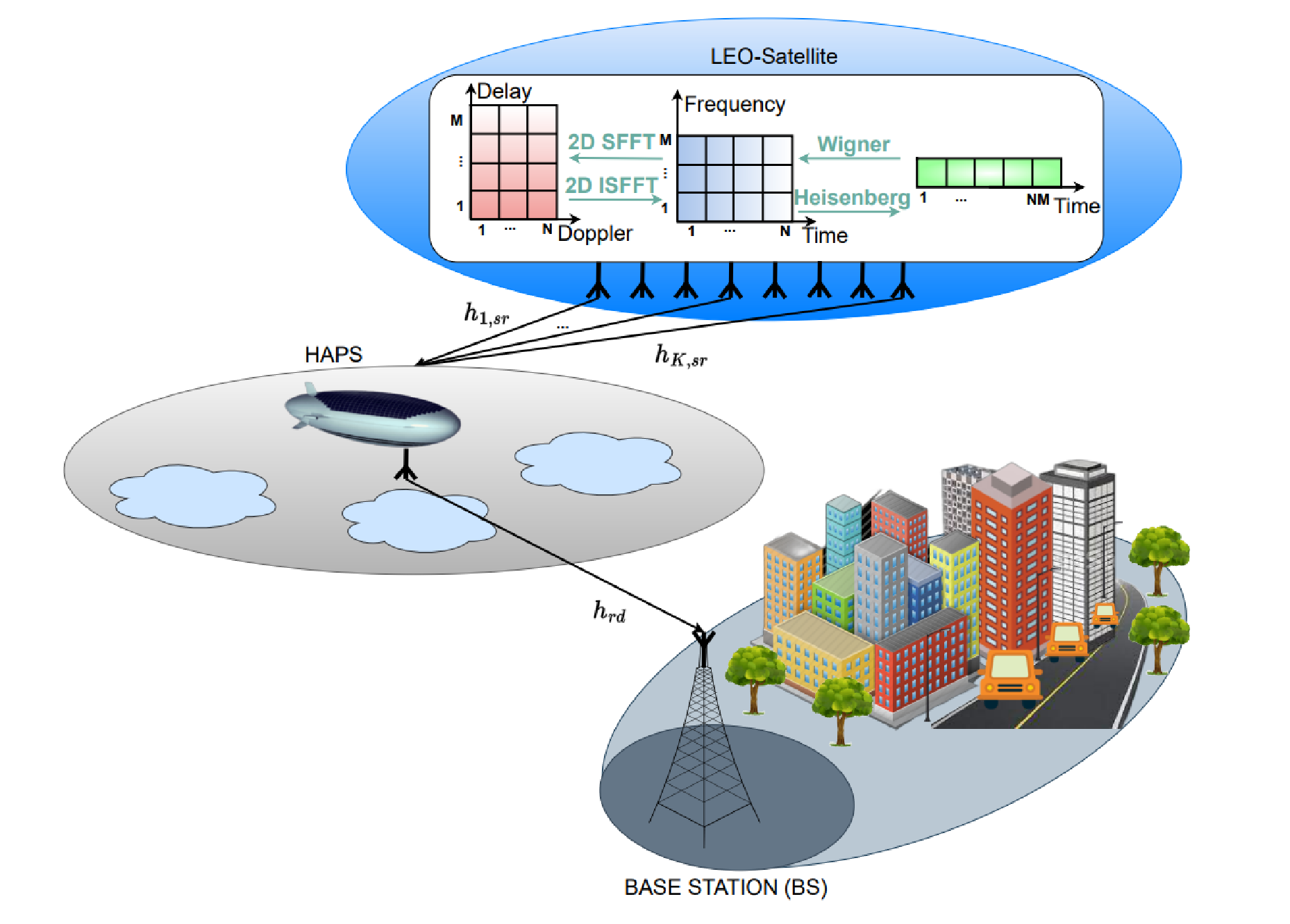}
    \caption{A downlink LEO-satellite network with HAPS relaying.}
    \label{fig:1}
\end{figure}

An OTFS waveform is utilized for both wireless communication links to remedy the Doppler effect in this system. The LEO-satellite carries out a wireless transmission to the HAPS using $N\times M$ information-bearing signals, { where $N$ and $M$ are the parameters used to partition Doppler and delay domains, respectively~\cite{ding2019robust}.} Under the OTFS scheme, the transmitted signals in the DD domain are denoted as $X[m,n]$, where $m \in \{0, \ldots, M-1\}$ and $n \in \{0, \ldots, N-1\}$. By performing inverse symplectic finite Fourier transform (ISFFT) as {depicted in Fig. \ref{fig:1} }, $X[m, n]$ can be mapped to the TF domain as
{
\begin{equation}
X_{tf}[l, k]=\frac{\sqrt{P_s}}{\sqrt{N M}} \sum_{n=0}^{N-1} \sum_{m=0}^{M-1} X[m, n] e^{j 2 \pi\left(\frac{n k}{N}-\frac{m l}{M}\right)},
\end{equation}
}where $P_s$ is transmit power of the LEO-satellite, $l \in \{0, \ldots, M-1\}$, and $k \in \{0, \ldots, N-1\}$. { The block circulant property of the channel matrix for the DD domain is assumed by utilizing an ideal bi-orthogonal pulse \cite{Delay-Doppler}}. $X_{tf}[l, k]$ is converted to the time domain (TD) signal from the TF domain using the Heisenberg transform. {The TD signal received by the HAPS can be expressed in the TF domain by utilizing the Wigner transform as}
{
\begin{equation}
Y_{tf}[l, k]=\frac{1}{\sqrt{d^{\alpha}}}H_{tf}[l, k] X_{tf}[l, k]+W_{tf}[l, k],
\end{equation}
}where $d$ is a distance between the LEO-satellite and the HAPS, $\alpha$ is a free space path-loss exponent, and $H_{tf}[l, k]=\iint h(\tau, \nu) e^{j 2 \pi \nu k T} e^{-j 2 \pi l \Delta f \tau} d \tau d v$. $\tau$ and $\nu$ denote delay and Doppler shift, respectively. The channel gain $h$ from the LEO-satellite to the HAPS is modeled with SR distribution. Note that $W_{tf}[l, k]$ is a complex Gaussian distributed noise in the TF domain. The TF domain signal $Y[m, n]$ is transformed to the DD domain signal by performing symplectic finite Fourier transform (SFFT)
{
\begin{equation}
    Y[m, n]=\frac{1}{\sqrt{N M}} \sum_{k=0}^{N-1} \sum_{l=0}^{M-1} Y_{tf}[l, k] e^{-j 2 \pi\left(\frac{k n}{N}-\frac{l m}{M}\right)} .
\end{equation}
}
$Y[m, n]$ can be expressed in the vectorized form as
{
\begin{equation}\label{eq:4}
\mathbf{y}=\sqrt{{\frac{P_s}{d^\alpha}}}  {\mathbf{H} \mathbf{x}} +\mathbf{w},
\end{equation}
}
where $\mathbf{x}$=$\left[\mathbf{x}_{0}^T \cdots \mathbf{x}_{n}^T \cdots \mathbf{x}_{N-1}^T\right]^T$. Note that  $\mathbf{H} \in \mathbb{C}^{{NM \times NM}}$ is a block circulant matrix and $\mathbf{x}_{n}^T$=$[x[n, 0] \cdots x[n, M-1]]^T$.
\subsection{Channel Model}
A wireless channel from an LEO-satellite to a HAPS is typically modeled with an SR distribution \cite{New_Model}, where the probability density function (PDF) of the instantaneous power is expressed for $v${-th} ($1 \leq v \leq K$) antenna of the LEO-satellite 
with ${ }_1 F_1(\cdot ; \cdot ;)$ as the confluent hypergeometric function
\begin{equation} 
\label{eq_1}
\begin{aligned}
&f_{h_{v}^2}(x)=\alpha_{v} \exp \left(-\beta_{v} x\right)_1 F_1\left(m_{v} ; 1 ; c_{v} x\right), \quad x \geq 0 \\
&\alpha_{v}   \triangleq \frac{\left(\frac{2 b_{0, v} m_{v}}{2 b_{0, v} m_{v}+\Omega_{v}}\right)^{m_{v}}}{2 b_{0, v}}, \quad
\beta_{v}  \triangleq \frac{1}{2 b_{0, v}}, \\
&c_{v}  \triangleq \frac{\Omega_{v}}{2 b_{0, v}\left(2 b_{0, v} m_{v}+\Omega_{v}\right)}.
\end{aligned}
\end{equation}
{$m_v, b_{0, v}$, and $\Omega_v$ are the Rayleigh and Nakagami distribution parameters. The lowpass-equivalent complex envelope of the stationary narrowband SR model between $v$-th transmit antenna of the LEO-satellite and the HAPS is obtained using the Rayleigh and Nakagami channel models.} $\Omega_v$ and $2b_{0, v}$ are the average power of the LOS and scatterer paths, respectively. $m_v$ represents a fading severity parameter. We assume that $m_v$ takes integer values for the sake of analytical tractability. Therefore, ${ }_1 F_1\left(m_{v} ; 1 ; c_{v} x\right)$ can be expressed as follows
\begin{equation}
\begin{aligned}
\label{F_1}
{ }_1 F_1\left(m_{v} ; 1 ; c_{v} x\right)&= \exp \left(c_{v} x\right) \\
& \times \sum_{k=0}^{m_{v}-1} \frac{(-1)^k\left(1-m_{v}\right) k\left(c_{v} x\right)^k}{(k !)^2}
\end{aligned}
\end{equation}
Using Eq.(\ref{F_1}), Eq.(\ref{eq_1}) can be expressed as follows 
\begin{equation}
\begin{aligned}
f_{h_{v}^2}(x)&=\sum_{k=0}^{m_{v}-1} \underbrace{\frac{(-1)^k\left(1-m_{v}\right)_k c_{v}^k}{(k !)^2}}_{\xi(k)} \\
&\times \alpha_{v} x^k \exp \left(-\left(\beta_{v}-c_{v}\right) x\right),
\end{aligned}
\end{equation}
where $(\cdot)_p$ is a Pochammer symbol. { Since the MRT scheme is utilized for the wireless communication link between the LEO-satellite and the HAPS, the sum of $K$ squared i.i.d. channel coefficients, $z \triangleq h_1^2+h_2^2+\cdots+h_K^2$, needs to be calculated.} The pdf of $z$ is expressed in closed-form as 
\begin{equation}
\begin{aligned}
&f_z(z)=\sum_{k_1=0}^{m_1-1} \cdots \sum_{k_K=0}^{m_1-1} \Xi(K) z^{\Lambda_1-1} \exp \left(-\left(\beta-c\right) z\right)\\
&\Xi(K) \triangleq \prod_{i=1}^K \xi\left(k_i\right) \alpha_1^K \prod_{j=1}^{K-1} B\left(\sum_{l=1}^j k_l+j, k_{j+1}+1\right)
\end{aligned}
\end{equation}
where $\Lambda_1 \triangleq \sum_{i=1}^K k_i+K$ and $B(\cdot,\cdot)$ is a beta function~\cite{Dual_Hop}.

\section{Outage Probability Analysis of OTFS}
A zero-forcing linear equalizer (ZF-LE) in the frequency domain (FD) is employed to eliminate the inter-symbol interference (ISI) of the signal in the receiver. By performing the ZF-LE over the received signal {at HAPS in Eq. \ref{eq:4},} we obtain
{
\begin{equation}
    \Theta_{sr} \mathbf{y}_{s r}=\sqrt{\frac{P_s}{d_{s r}^{\alpha}}} \mathbf{x}+\Theta_{sr} \mathbf{w}_{s r},
\end{equation}
}where $\Theta_{sr}=\left(\mathbf{F}_N^H \otimes \mathbf{F}_M\right) \mathbf{D}_{s r}^{-1}\left(\mathbf{F}_N \otimes \mathbf{F}_M^H\right)$ { \cite{equalization}}. Note that $(\cdot)^{T}$ and $(\cdot)^{H}$ denote the transpose and hermitian operations of a matrix, respectively. $\mathbf{F}_{n}$ is a $n\times n$ discrete Fourier transform matrix, {$D_{sr} = \sum_{i=1}^{K} D_{i,sr}$, and a diagonal matrix $\mathbf{D}_{i, s r}$ whose $(kM+l+1)${-th} diagonal element can be expressed as}
\begin{equation}
D_{i, s r}^{k, l}=\sum_{n=0}^{N-1} \sum_{m=0}^{M-1} a_{i, s r, n}^{m, 1} e^{j 2 \pi \frac{l m}{M}} e^{-j 2 \pi \frac{k n}{N}},
\end{equation}
where $a^{m,1}_{i,s r, n}$ is the element located in the position $(nM + m + 1, 1)$ of {the DD domain channel matrix between $i${-th} transmit antenna and the receiver}. The covariance matrix of $\Theta \mathbf{w}_{s r}$, which is denoted by $\phi_{s r}$, is calculated to derive the SNR expression of the received signal in the HAPS as
\begin{equation}
\begin{aligned}
    \label{phi_sr}
    \phi_{s r} &
    =\mathrm{E}\left[ (\Theta \mathbf{w}_{s r})(\Theta \mathbf{w}_{s r})^H \right] \\ 
    & =\mathrm{E}\left[\left(\mathbf{F}_N^H \otimes \mathbf{F}_M\right) \mathbf{D}_{s r}^{-1} \mathbf{w}_{s r}\mathbf{w}_{s r}^H \mathbf{D}_{s r}^{-H} \left(\mathbf{F}_N \otimes \mathbf{F}_M^H\right)\right] \\    
    & =\left(\mathbf{F}_N^H \otimes \mathbf{F}_M\right) \mathbf{D}_{s r}^{-1}\mathbf{D}_{s r}^{-H}\left(\mathbf{F}_N \otimes \mathbf{F}_M^H\right),
\end{aligned}
\end{equation}
where $\otimes$  and $\mathrm{E}\left[\cdot\right]$ represent the Kronecker product and the expected value expression, respectively. Note that $\left(\mathbf{F}_N \otimes \mathbf{F}_M^H\right) \mathbf{w}_{s r}$ is subject to a complex normal distribution denoted by $ C N\left(0, \mathbf{I}_{N M}\right)$. Hence, $\phi_{s r}$ can be expressed as
\begin{equation}
\begin{aligned}\label{eq.12}
    \phi_{s r} & =\frac{1}{N M} \operatorname{Tr}\left[\left(\mathbf{F}_N^H \otimes \mathbf{F}_M\right) \mathbf{D}_{s r}^{-1}\mathbf{D}_{s r}^{-H}\left(\mathbf{F}_N \otimes \mathbf{F}_M^H\right)\right] \\
    & =\frac{1}{N M} \sum_{\hat{k}=0}^{N-1} \sum_{\hat{l}=0}^{M-1} || D_{ s r}^{\hat{k}, \hat{l}} || ^{-2},
\end{aligned}
\end{equation}
where $\mathrm{Tr}\left[\cdot\right]$ denotes the trace of the matrix. After the HAPS decodes the received message, the decoded message is re-encoded and forwarded from the HAPS to the BS as $\mathbf{\tilde{x}}$. Hence, $y_{r d}[k, l]$ can be expressed in the vectorized form as
{
\begin{equation}
\mathbf{y}_{r d}=\sqrt{{\frac{P_s}{d_{r d}^\alpha}}}  {\mathbf{H}_{r d} \mathbf{\tilde{x}}} +\mathbf{w}_{r d}.
\end{equation}
}By performing ZF-LE over the received signal from the HAPS to the BS, we obtain
{
\begin{equation}
    \Theta_{r d} \mathbf{y}_{r d}=\sqrt{\frac{P_s}{d_{r d}^{\alpha}}} \mathbf{\tilde{x}}+\Theta_{r d} \mathbf{w}_{r d},
\end{equation}
}where $\Theta_{r d}=\left(\mathbf{F}_N^H \otimes \mathbf{F}_M\right) \mathbf{D}_{r d}^{-1}\left(\mathbf{F}_N \otimes \mathbf{F}_M^H\right)$. Using Eq. \ref{phi_sr}, $\phi_{r d}$ can be expressed as
\begin{equation}
\begin{aligned}
    \phi_{r d} & =\frac{1}{N M} \operatorname{Tr}\left[\left(\mathbf{F}_N^H \otimes \mathbf{F}_M\right) \mathbf{D}_{r d}^{-1}\mathbf{D}_{r d}^{-H}\left(\mathbf{F}_N \otimes \mathbf{F}_M^H\right)\right] \\
    & =\frac{1}{N M} \sum_{\hat{k}=0}^{N-1} \sum_{\hat{l}=0}^{M-1} | D_{r d}^{\hat{k}, \hat{l}} | ^{-2}.
\end{aligned}
\end{equation}
Hence, the SNR of the received signal at the HAPS ($q = sr$) and the BS ($q = rd$) is given by 
\begin{equation}
    \gamma_{q}=\frac{P_s}{\sigma_{q}^2 \phi_{q} d_{q}^{\alpha}}, \quad q \in [sr, rd].
\end{equation}
 Note that, the transmitted signal from $i${-th} antenna to the HAPS is multiplied with the following weight factor
\begin{equation}
    w_i^{{\tilde k, \tilde l}} = \frac{({D_{i,sr}^{\tilde k,\tilde l})^H}}{\sqrt{\sum_{k=1}^{K}  \left| { D^{\tilde k, \tilde l}_{k,sr}}\right|^2}}.
\end{equation}
{Therefore, $\phi_{sr}$ in Eq. \ref{eq.12} can be expressed as}
\begin{equation}
    {\phi_{sr} = \frac{1}{NM}\sum^{N-1}_{\tilde k=0}\sum^{M-1}_{\tilde l=0} \left| \sum_{k=1}^{K} \left|D^{\tilde k, \tilde l}_{k,sr}  w_k^{\tilde k,\tilde l}\right| \right|^{-2}.}
\end{equation}
{After some algebraic operations, $\phi_{sr}$ becomes}
\begin{equation}
    \phi_{sr} = \frac{1}{NM}{\sum^{N-1}_{\tilde k=0}\sum^{M-1}_{\tilde l=0}} \left| \frac{1}{\sum_{k=1}^{K} \left|D^{{\tilde k, \tilde l}}_{k,sr}\right|^2}\right|.
\end{equation}
Using the central limit theorem, $\phi_{sr}$ can be modeled with a Gaussian distribution for higher values of $N$ and $M$. The expected value and the variance of $\phi_{sr}$ can be expressed as
\begin{equation}
    \mathbb{E}[\phi_{sr}] = \mathbb{E}\left[\frac{1}{{\rho}}\right] = \int_{0}^{\infty} \frac{f_{\rho} ({\rho})}{{\rho}} d{\rho},
\end{equation}
\begin{equation}
    \mathbb{V}[\phi_{sr}] = \frac{1}{NM}\mathbb{V}\left[\frac{1}{{\rho}}\right] = \frac{1}{NM} \left(\mathbb{E}\left[\frac{1}{{\rho}^2}\right] -\mathbb{E}\left[\frac{1}{{\rho}}\right]^2\right),
\end{equation}
 where $\sum_{k=1}^{K} |h_{k,sr}|^2$ is called ${\rho}$ for simplicity. The closed-form expressions of  $\mathbb{E}\left[\frac{1}{{\rho}^n}\right]$ is derived for {$K$} transmit antennas as follows~\cite{Table_of_Integral}
\begin{equation}
\mathbb{E} \left[\frac{1}{{\rho}^n}\right] = \sum_{k_1=0}^{m_1-1} \cdots  \sum_{k_K=0}^{m_1-1}\Xi_1(K)\frac{\Gamma(\lambda_1-n)}{\left(\beta-c\right)^{\lambda_1-n}}.
\end{equation}
For the second link, the model proposed in \cite{Outage} is used. Since $ D_{ r d}^{\hat{k}, \hat{l}}$ is Nakagami-$m$ distributed, $\left|D_{r d}^{\tilde{k}, \tilde{l}}\right|^{-2} \sim \operatorname{IG}\left(\alpha_{IG}, \beta_{IG}\right)$ is subject to the inverse gamma distrubution. Furthermore, $\phi_{rd}$ is approximated to a gamma distribution $\phi_{rd} \sim \operatorname{G}\left(\alpha_{G}, \beta_{G}\right)$, where the mean and variance of the gamma distribution is set to the mean and variance of $\phi_{rd}$,
\begin{equation}
    \begin{aligned}
        \alpha_{G} = NM (\alpha_{IG}-2),
    \end{aligned}
\end{equation}
\begin{equation}
    \begin{aligned}
        \beta_{G} = NM \frac{(\alpha_{IG}-1)(\alpha_{IG}-2)}{\beta_{IG}}.
    \end{aligned}
\end{equation}
The {end-to-end} {OP} of the dual-hop communication system can be calculated as follows
\begin{equation}\label{eq:Pout}
\begin{aligned}
    \left(1-P_{out}\right) =  \left(1-P_{out1}\right) +  \left(1-P_{out2}\right)
\end{aligned}
\end{equation}
{where,} $P_{out1}$ and $P_{out2}$ define the {OP} of dual-hop links. By making trivial calculations in \eqref{eq:Pout}, the end-to-end OP is expressed with the following equation
\begin{equation}\label{eq:Pout2}
\begin{aligned}
     P_{out} = 1-\mathbf{Pr}(\gamma_{s r}>\gamma_{th})\mathbf{Pr}(\gamma_{r d}>\gamma_{th}),
\end{aligned}
\end{equation}
where $\gamma_{th}$ denotes the threshold SNR of the proposed system, $\mathbf{Pr}(\cdot)$ denotes the probability expression. After some mathematical calculations, the {OP} for the first link is derived as
\begin{equation}
\begin{aligned}\label{eq:q_func}
    \mathbf{Pr}\left(\frac{P_s}{\sigma_{sr}^2 \phi_{sr} d_{sr}^{\alpha}}\leq \gamma_{th}\right) 
    = Q\left(   \frac{\frac{P_s}{\sigma_{sr}^2 d_{sr}^{\alpha} \gamma_{th}} - \mathbb{E}[\phi_{sr}]}{\sqrt{\mathbb{V}[\phi_{sr}]}} \right)
\end{aligned}
\end{equation}
{where $Q(\cdot)$ is the tail distribution function of the standard normal distribution.}
For the second link, the {OP} of the Nakagami-$m$ channel is calculated using the cdf of the Gamma distribution 
\begin{equation}
\begin{aligned}\label{eq:gamma_func}
    \mathbf{Pr}\left(\frac{P_s}{\sigma_{rd}^2 \phi_{rd} d_{rd}^{\alpha}}\leq \gamma_{th}\right) 
    =1-\frac{ \Gamma\left( \alpha_{G},\beta_G\frac{P_S}{\sigma_{rd}^2 d_{rd}^{\alpha} \gamma_{th}} \right)} {\Gamma\left( \alpha_G\right)}
\end{aligned}
\end{equation}
where $\Gamma\left(\cdot\right)$ represents the gamma function and $\Gamma\left(\cdot,\cdot\right)$ denotes the incomplete gamma function.
Finally, $P_{out}$ can be calculated using Eqs. \eqref{eq:q_func} and \eqref{eq:gamma_func} into Eq. \eqref{eq:Pout} to obtain Eq. \eqref{eq:end2end}.
\begin{table*}[!t]
\centering
\begin{equation}\label{eq:end2end}
        P_{out} = Q\left(\frac{\frac{P_s}{\sigma_{sr}^2 d_{sr}^{\alpha} \gamma_{th}} - \mathbb{E}[\phi_{sr}]}{\sqrt{\mathbb{V}[\phi_{sr}]}} \right) + \frac{\Gamma(\alpha_{G})- \Gamma\left( \alpha_{G},\beta_G\frac{P_S}{\sigma_{rd}^2 d_{rd}^{\alpha} \gamma_{th}} \right)} {\Gamma\left( \alpha_G\right)} 
    - Q\left(\frac{\frac{P_s}{\sigma_{sr}^2 d_{sr}^{\alpha} \gamma_{th}} - \mathbb{E}[\phi_{sr}]}{\sqrt{\mathbb{V}[\phi_{sr}]}} \right) \frac{\Gamma(\alpha_{G})- \Gamma\left(\alpha_{G},\beta_G\frac{P_S}{\sigma_{rd}^2 d_{rd}^{\alpha} \gamma_{th}} \right)} {\Gamma\left( \alpha_G\right)}
\end{equation}
\hrule
\end{table*}

\section{Simulation Results} 
This section presents the analytical and simulation results of the {OP} for the proposed OTFS waveform under different channel conditions. Simulations are constructed and run in MATLAB for two different channel conditions. The values of the channel parameters are given in Table \ref{tab:1} for both frequent heavy shadowing (FHS) \cite{New_Model} and Karasawa\cite{karasawa}. $\gamma_{th}$ is considered $0$ dB for all links. The number of Monte-Carlo simulations is set to $10^{7}$.
\begin{table}[!t]  
\caption{Channel Parameters}
\label{tab:1}
\centering 
\begin{tabular}{l c c c} 
\hline\hline 
 Channel Condition & $m$ & $b$ & $\Omega$
\\ [0.5ex]  
\hline   
 Frequent Heavy Shadowing & 1 & 0.063 & $7\times10^{-4}$\\  [1ex] 
Data set of Karasawa et al. & 2 & 0.0158 & 0.123 \\[1ex]   
\hline 
\end{tabular}  
\end{table}  
\begin{figure}[t]
\centering
\subfloat[]{\label{fig:pdf:a} 
\includegraphics[width=0.48\columnwidth]{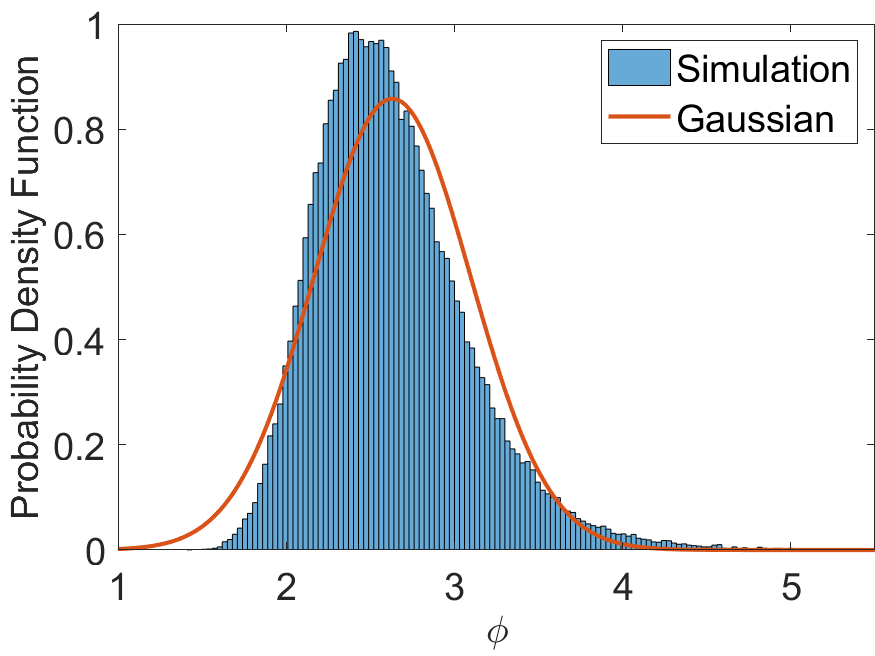}}
\subfloat[]{\label{fig:pdf:b} 
\includegraphics[width=0.48\columnwidth]{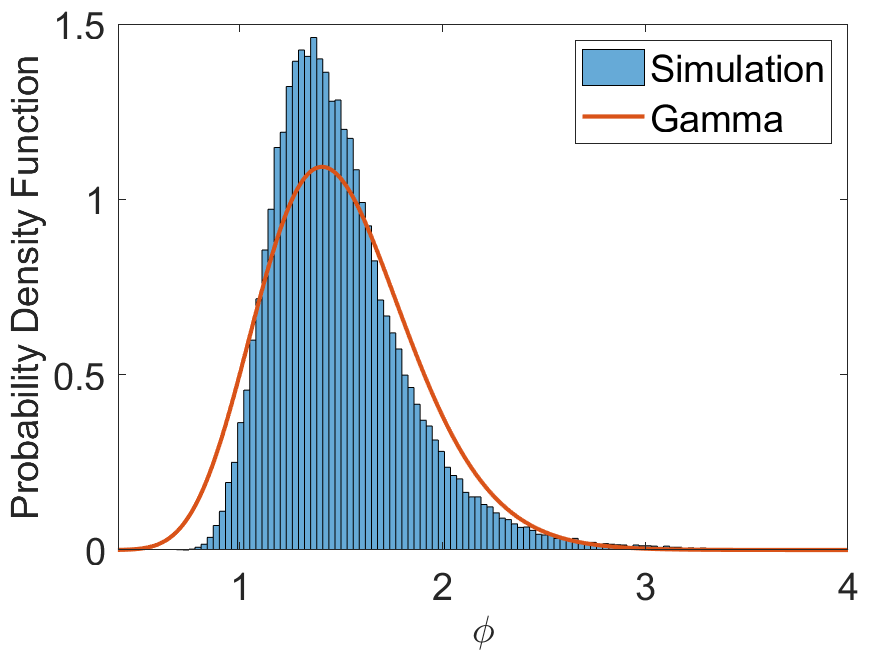}}\\
\subfloat[]{\label{fig:pdf:c} 
\includegraphics[width=0.48\columnwidth]{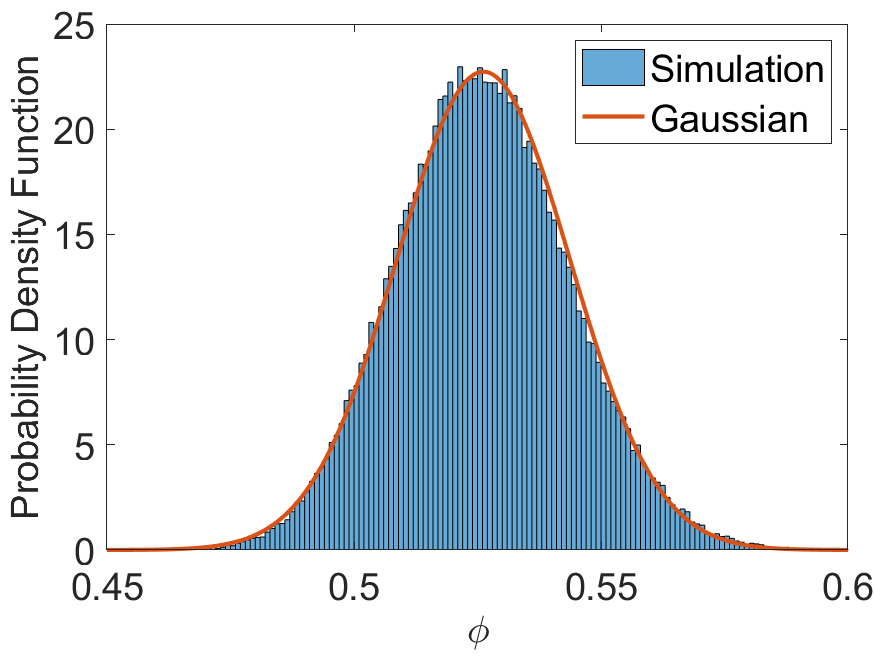}}
\subfloat[]{\label{fig:pdf:d} 
\includegraphics[width=0.48\columnwidth]{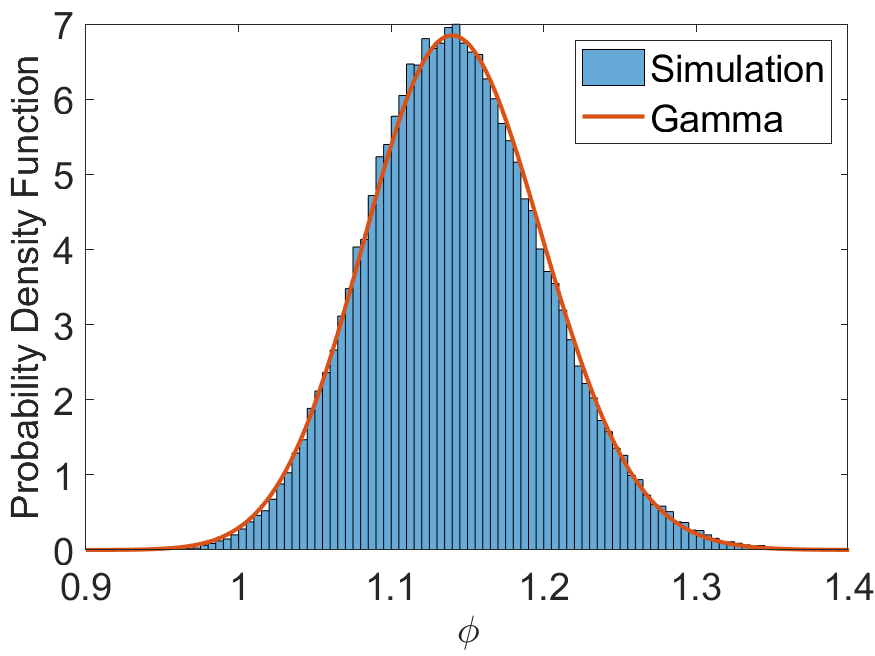}}
\caption{The pdfs and histograms for (a) {$K$}=4, $N$=$M$=4; (b) $N$=$M$=4, $m$=3; (c) {$K$}=16, $N$=$M$=8; (d) $N$=$M$=8, $m$=8}
\label{fig:pdfs}
\end{figure}

{ Fig. \ref{fig:pdfs} demonstrates the effect of OTFS orders and the number of antennas on the accuracy of the Gaussian and Gamma distribution approximations in the proposed analytical model. The results clearly illustrate that as the number of antennas and OTFS order increase, the Gaussian approximation aligns more closely with the original distribution, while the Gamma approximation better fits the original distribution with increasing $m$ parameter and OTFS order. We employed two distinct metrics, namely the normalized mean square error (NMSE) \cite{Outage} and the Kullback–Leibler divergence (KL-divergence), to assess the fitness of the proposed analytical model. Firstly, the NMSE is reduced from $0.0550$ (Fig. \ref{fig:pdf:a}) to $0.0021$ (Fig. \ref{fig:pdf:c}) for the first link and $0.0692$ (Fig. \ref{fig:pdf:b}) to $0.0012$ (Fig. \ref{fig:pdf:d}) for the second link. Secondly, KL-divergence decreased from $0.0863$ (Fig. \ref{fig:pdf:a}) to $0.0027$ (Fig. \ref{fig:pdf:b}) for the first link, and decreased from $0.0816$ (Fig. \ref{fig:pdf:c}) to $0.0027$ (Fig. \ref{fig:pdf:d}) for the second link. These results indicate that the Gaussian and Gamma distribution approximations become more accurate as $N$, $M$, $m$, and $K$ increase.}

\begin{figure*}[t]
\centering
\subfloat[]{\label{fig:Result:a} 
\includegraphics[width=0.32\linewidth]{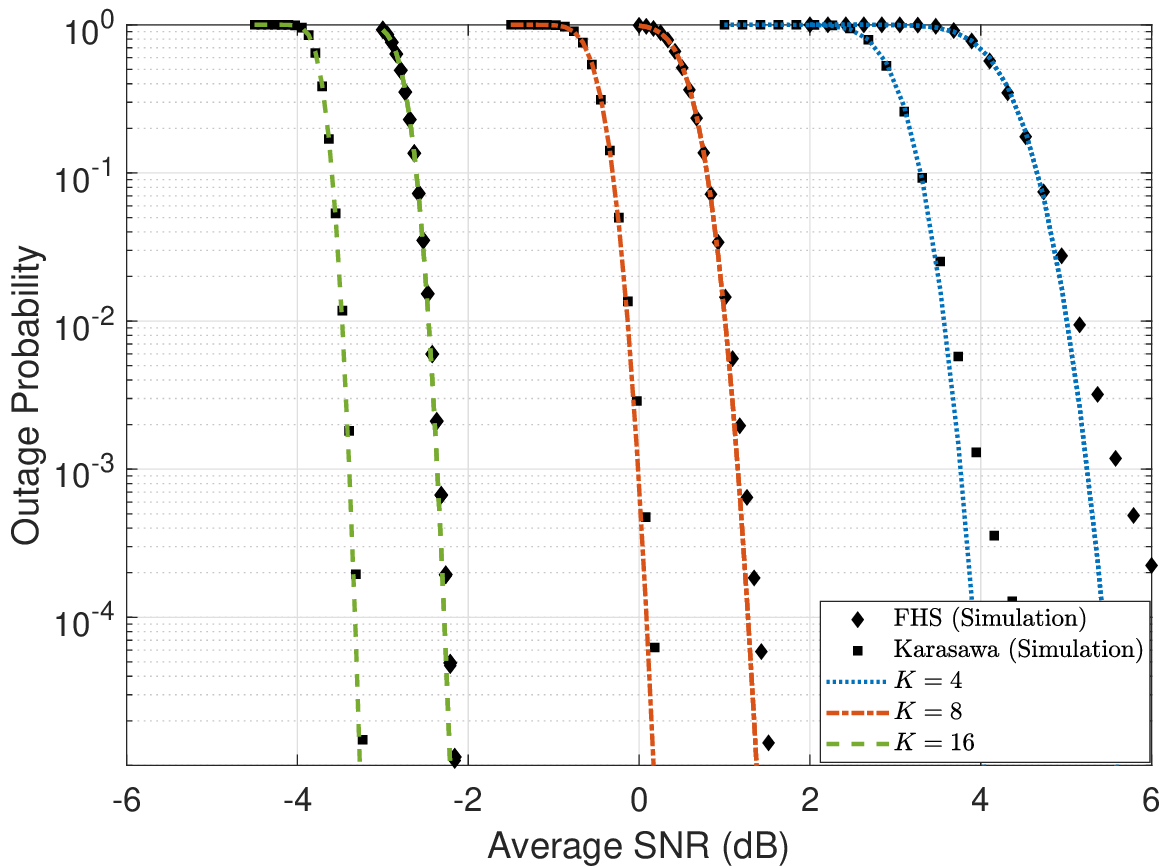}}
\subfloat[]{\label{fig:Result:b} 
\includegraphics[width=0.32\linewidth]{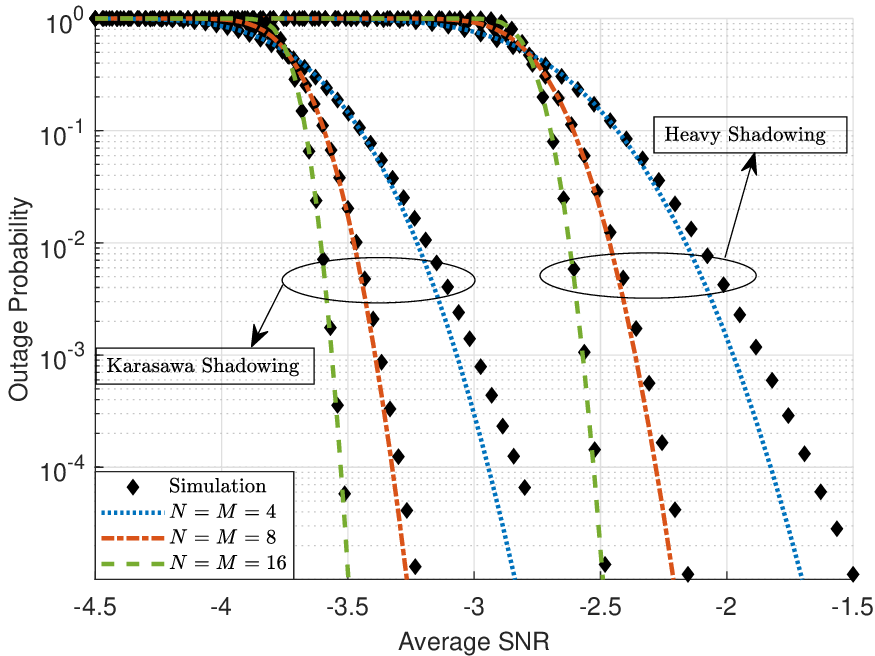}}
\subfloat[]{\label{fig:Result:c} 
\includegraphics[width=0.32\linewidth]{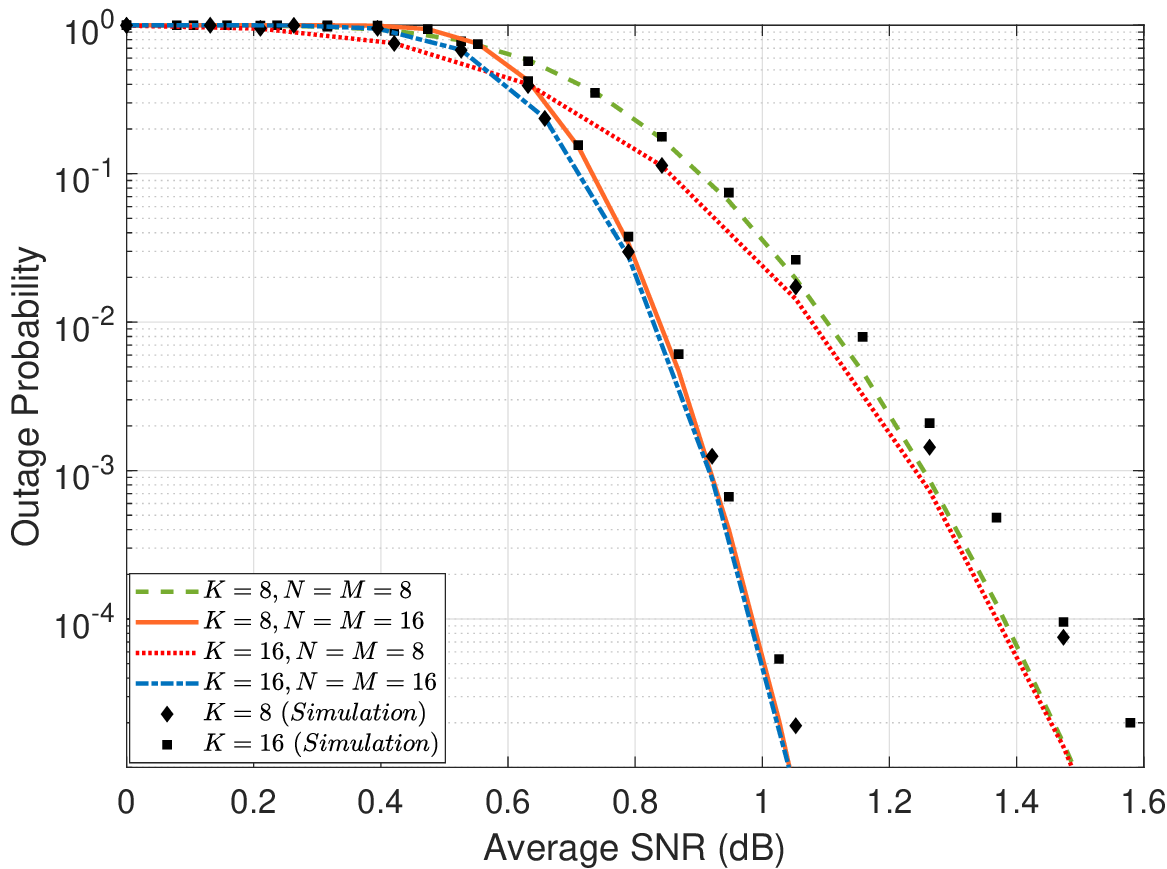}}
\caption{The {OP} versus the average SNR for the (a) first link with $N$=$M$=8; (b) first link with {$K$} = 8; (c) end-to-end system.}
\end{figure*}

Fig. \ref{fig:Result:a} illustrates the {OP} as a function of the average SNR for the LEO-satellite to HAPS link under FHS and Karasawa channel conditions. It is observed that the {OP} decreases across all cases as the average SNR increases from $-6$ to $6$ dB. The comparison between analytical and simulation results reveals that the accuracy improves as the number of antennas increases. Particularly, the highest level of accuracy is attained when $K = 16$. It is worth noting that, for the same number of antennas, the Karasawa channel condition consistently outperforms the Frequent Heavy shadowing condition as anticipated.

The {OP} with respect to the average SNR for FHS and Karasawa channel, which is used to model the LEO-satellite to HAPS link is depicted in Fig. \ref{fig:Result:b}. The {OP} decreases as the average SNR increases from $-4.5$ to $-1.5$ dB. The significance of the channel conditions on the system performance is clearly observed for a fixed number of antennas. Additionally, increasing the OTFS order contributes to improved diversity order and steeper slopes in the {OP results.} The proposed analytical model achieves the closest alignment with the simulation results when the OTFS order is set to $N=M=16$. This suggests that the model accurately captures the behaviour of the system under these specific conditions, validating its effectiveness.

Fig. \ref{fig:Result:c} shows the simulation and analytical results of the {OP} for the end-to-end dual-hop system, including two wireless links from the satellite to the HAPS with heavy shadowing and another from the HAPS to the BS with $m=8$. The {OP} of the HAPS to the BS link becomes a dominating factor in determining the system {OP} for given channel conditions. Therefore, increasing the OTFS order is preferable instead of using a higher number of antennas in the first link. The proposed mathematical framework is able to model arbitrary values of $N, M$, and $K$. However, for verifying the mathematical framework, we utilize $N, M$, and $K$ values up to $16$ due to the computational complexity in the simulation experiments.
\section{Conclusion}
This study investigated how the OTFS waveform can be used in multi-antenna transmission with DF relaying, specifically in the context of LEO-satellite channel conditions. A closed-form expression for the {OP} of the ZF equalization is derived when the MRT scheme is utilized. Our results, without loss of generality, demonstrate that increasing information-bearing signals and the number of antennas leads to lower {OP}. In addition, the {OP} approximation becomes more accurate as the OTFS order and the number of transmit antennas increase. The design of effective and trustworthy communication systems for LEO-satellite channels will be inspired by these findings in a practical way. The potential of OTFS in maximum ratio combining and its practical modelling could be investigated as part of future work.

\section*{Acknowledgement}
This study has been carried out through the research vision of the THULAB project run at the Informatics and Information Security Research Center (B{\.I}LGEM) of The Scientific and Technological Research Council of Türkiye (T{\"U}B{\.I}TAK).

This work has also received funding from the AIMS5.0 project. AIMS5.0 has been accepted for funding within the Key Digital Technologies Joint Undertaking (KDT JU), a public-private partnership in collaboration with the HORIZON Framework Programme and the national Authorities under grant agreement number 101112089.

\bibliographystyle{IEEEtran}
\bibliography{main.bib}

\vfill

\end{document}